\documentclass[conference]{IEEEtran}
\IEEEoverridecommandlockouts
\usepackage{cite}
\usepackage{amsmath,amssymb,amsfonts}
\usepackage{graphicx}
\usepackage{algorithm}
\usepackage{algorithmic}
\usepackage{subcaption}
\usepackage{paralist}
\usepackage{xcolor}
\usepackage{url}
\usepackage{microtype}
\usepackage{orcidlink}
\hypersetup{hidelinks}

\graphicspath{{ICDE/}}

\renewcommand\leq\leqslant
\renewcommand\geq\geqslant

\providecommand{\Description}[1]{}
\providecommand{\Small}{\ifmmode\else\footnotesize\fi}

\newcommand{\newpart}[1]{{\color{black}#1}}
\newcommand{\sylli}[1]{{\color{black}#1}}
\newenvironment{newpartenv}{\color{black}}{\ignorespacesafterend}

\makeatletter\g@addto@macro\normalsize{%
  \setlength{\abovedisplayskip}{5pt plus 2pt minus 2pt}\setlength{\belowdisplayskip}{5pt plus 2pt minus 2pt}%
  \setlength{\abovedisplayshortskip}{2pt plus 1pt}\setlength{\belowdisplayshortskip}{3pt plus 1pt minus 1pt}}\makeatother
\makeatletter
\def\section{\@startsection{section}{1}{\z@}{1.0ex plus 1ex minus 0.5ex}{0.5ex plus 0.5ex minus 0ex}{\normalfont\normalsize\centering\scshape}}
\def\subsection{\@startsection{subsection}{2}{\z@}{1.0ex plus 1ex minus 0.5ex}{0.5ex plus 0.5ex minus 0ex}{\normalfont\normalsize\itshape}}
\makeatother
\begin{document}

\title{Toward a Scalable and Adaptive Evaluation Measure \\ for Time Series Anomaly Detection}

\author{
\IEEEauthorblockN{Emmanouil Sylligardos\,\orcidlink{0000-0001-5539-3209}\IEEEauthorrefmark{1},
Mingyi Huang\,\orcidlink{0009-0003-7287-3417}\IEEEauthorrefmark{2},
John Paparrizos\,\orcidlink{0000-0002-7592-748X}\IEEEauthorrefmark{2}\IEEEauthorrefmark{3},
Pierre Senellart\,\orcidlink{0000-0002-7909-5369}\IEEEauthorrefmark{1},
Paul Boniol\,\orcidlink{0000-0001-8516-0123}\IEEEauthorrefmark{1}}
\IEEEauthorblockA{\IEEEauthorrefmark{1}DI ENS, ENS, PSL Univ., CNRS, Inria, Paris, France}
\IEEEauthorblockA{\IEEEauthorrefmark{2}The Ohio State Univ., Columbus, OH, USA \quad \IEEEauthorrefmark{3}Aristotle Univ. Thessaloniki, Thessaloniki, Greece}
\IEEEauthorblockA{emmanouil.sylligardos@ens.fr, huang.5749@osu.edu, paparrizos.1@osu.edu, pierre@senellart.com, paul.boniol@inria.fr}
}

\maketitle

\begin{abstract}
  \newpart{Time Series Anomaly Detection (TSAD) is a fundamental task for time series analytics. Unlike other domains, where anomaly detection mainly focuses on point-based anomalies, TSAD must also handle range-based anomalies. Nevertheless, TSAD methods are commonly evaluated with measures inherited from classification, such as Precision and Recall, which require a threshold, or with threshold-independent ones, such as AUC, which ignore the temporal aspect of anomalies. To address these issues, the Volume Under the Surface (VUS) measure scores detectors across all thresholds and a range of buffer lengths and has since been widely adopted for TSAD. However, VUS has two limitations. First, VUS is computationally expensive, requiring 7 hours and 38 minutes to evaluate 32 detectors on the 870 TSB-AD time series. Second, VUS applies the same buffer lengths to every anomaly and detector, which can yield over-optimistic scores that hinder the discrimination between accurate and mediocre detectors. To address these issues, we first introduce \textbf{FastVUS}, which computes VUS exactly and on average $76\times$ faster by (i) jointly computing all buffers in a parallel manner, (ii) pruning irrelevant buffers efficiently for the existence reward computation\sylli{, (iii) casting the confusion matrix as a single reduced matrix multiplication, (iv) computing VUS for every buffer length in one pass, and (v) discarding non-anomalous regions with a global mask}. We also develop \textbf{FastVUS-GPU}, which \sylli{recasts every step as batched tensor operations and} reaches an average $304\times$ speedup with a maximum error of $5\times10^{-3}$. Building on top of FastVUS, we then introduce \textbf{AutoVUS}, an adaptive extension that compares each detection against a random baseline and selects the buffer length that maximizes the gain over it. Our case studies show that AutoVUS reduces over-optimistic scores and better separates accurate from mediocre detectors. We release FastVUS and AutoVUS as open source to promote time-aware evaluation at scale.}
\looseness=-1
\end{abstract}

\begin{IEEEkeywords}
Time series, anomaly detection, evaluation measures
\end{IEEEkeywords}


\section{Introduction}
\label{sec:intro}

\begin{figure}[!t]
    \centering
    \begin{subfigure}[b]{\linewidth}
        \centering
        \includegraphics[width=\linewidth]{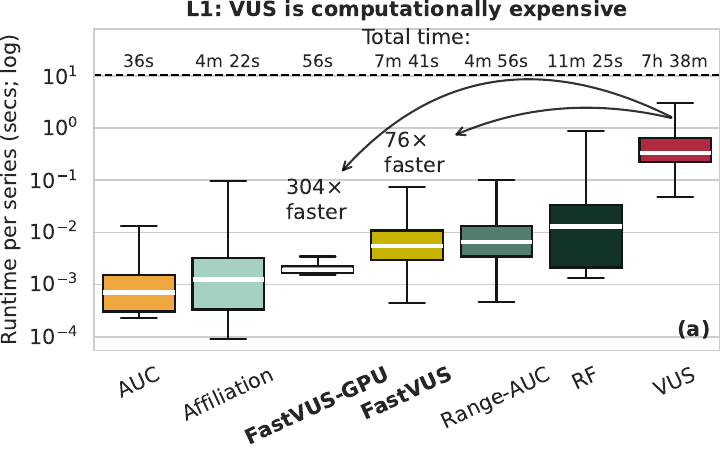}
        \vspace{-0.2cm}
        \phantomcaption
        \label{fig:intro_runtime}
    \end{subfigure}
    \begin{subfigure}[b]{\linewidth}
        \centering
        \includegraphics[width=\linewidth]{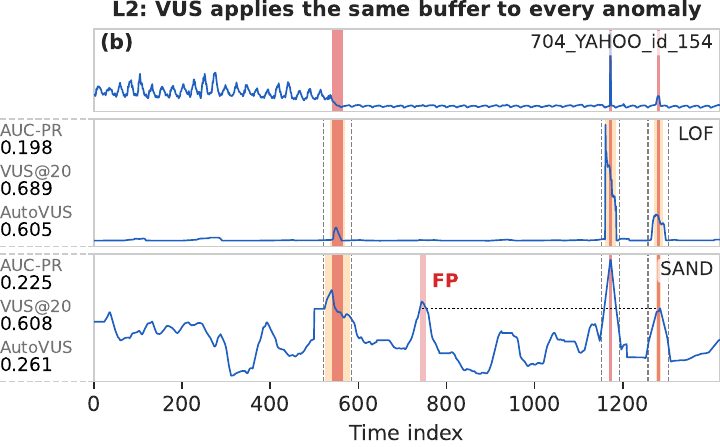}
        \phantomcaption
        \label{fig:intro_adaptive}
    \end{subfigure}
    \vspace{-0.7cm}
    \caption{\sylli{Our two main contributions. (a) Runtime per series and total time on TSB-AD. (b) An example series with two scores. Red marks the labeled anomalies, dashed lines the fixed VUS buffer, orange the buffers \textbf{AutoVUS} selects, \sylli{and light red a false positive (FP)}. VUS@20 is VUS-PR with $L=20$.}}
    \label{fig:intro}
\end{figure}

Time series data are pervasive across domains~\cite{Palpanas2019, liu2023amir}, from medical diagnostics~\cite{sylligardos2022deep} to environmental monitoring~\cite{goddard2003geospatial}. In all these domains, Time Series Anomaly Detection (TSAD), is a fundamental task~\cite{page1957problems, fox1972outliers} that flags faults, failures, and other critical events~\cite{barnett1994outliers, DBLP:conf/vldb/SubramaniamPPKG06, DBLP:conf/icdm/YehZUBDDSMK16}. Moreover, as the volume of time series grows, so does the demand for accurate detectors, and the community has responded with many detection methods~\cite{blazquez2021review, paparrizosTSB2022, schmidlBenchmark2022, liu2024elephant, paparrizos2022volume, jacob2020exathlon, Kim2021TowardsAR, 7424283, wu2020current}. 
Evaluation measures are responsible for assessing the quality of detectors. 
As a result, progress in TSAD is only as trustworthy as the measure that picks the best detector.

Yet popular evaluation measures were not built for time series, but inherited from classification, comprising threshold-dependent measures such as Precision, Recall, and the F-score, together with threshold-independent measures such as the Area Under the ROC or Precision-Recall curve (AUC)~\cite{fawcett2006introduction, 10.1145/1143844.1143874}. These served the community for decades, but a growing line of work argues that they are inadequate for TSAD~\cite{tatbul_precision_2018, huet2022local, salles2024softed, paparrizos2022volume}. This inadequacy lies in how these measures treat time, scoring each point independently, i.e., thresholding the anomaly score to mark each point as an anomaly or not~\cite{paparrizos2022volume}. A detection that fires shortly before or after a labeled anomaly points to the same event and is nearly correct, but under point-wise scoring it counts as a miss. This rigidity is costly in two situations common in TSAD: (a) annotators disagree on where an anomaly begins and ends, and (b) many detectors, such as forecasters and autoencoders, signal an anomaly slightly before or after it fully unfolds~\cite{blazquez2021review}.

To address these shortcomings, Tatbul et al.~\cite{tatbul_precision_2018} proposed range-based Precision, Recall, and F-score, which extend the traditional definitions to consider whether an anomaly is detected, how many of its points are detected, and which part. Range-AUC~\cite{paparrizos2022volume} and SoftED~\cite{salles2024softed} further add a \textit{buffer}, a tolerance region around each anomaly boundary, so that detections close to the boundary are rewarded rather than treated as misses. These measures, however, require fixing parameters in advance, such as the buffer length, and detector scores depend on that choice~\cite{paparrizos2022volume}. The Volume Under the Surface (VUS) measure~\cite{paparrizos2022volume} avoids fixing either a threshold or a buffer length by integrating performance over all thresholds and all buffer lengths up to a maximum $L$, producing one score robust to both threshold selection and boundary imprecision. VUS has since become a de facto choice for evaluating TSAD~\cite{chen2023imdiffusion, yang2023dcdetector, zhong2025simad, si2024timeseriesbench, goswami2024moment, ma2024survey}, \sylli{but still, VUS has two limitations.}

\noindent\sylli{\textbf{Limitation 1 (L1).} VUS is computationally expensive, as integrating over all thresholds and buffer lengths has} complexity $O(N_\theta N_b n)$~\cite{boniol2025vus}, where $N_\theta$ is the number of thresholds, $N_b$ the number of buffer lengths, and $n$ the length of the series. \sylli{VUS does not scale} to the large benchmarks and repeated evaluations that modern TSAD demands. For example, in a recent survey, some VUS results could not be reported because of the ``excessive computation time'' they required~\cite{ma2024survey}, and in our experiments, computing VUS for 32 detectors on the 870 time series of the TSB-AD benchmark~\cite{liu2024elephant} takes 7 hours and 38 minutes \sylli{(Figure~\ref{fig:intro_runtime})}. \newpart{A practitioner is thus left to choose between a measure that respects the temporal structure of time series but does not scale, such as VUS, and one that scales but was designed for isolated points, such as AUC.}

\newpart{\noindent\sylli{\textbf{Limitation 2 (L2).} VUS applies the same buffer lengths to every anomaly and detector. Averaging over all buffer lengths up to $L$ makes VUS less sensitive to any single one, but does not adapt the buffer to the tolerance each detection needs.} This uniform treatment can be too permissive. For example, a short anomaly may receive a buffer much longer than the event itself, and increasing the buffer can then raise the score by inflating the chance-level reward rather than by showing that the detector better localizes the event. Figure~\ref{fig:intro_adaptive} shows a representative example, in which LOF stays flat except near the three labeled anomalies, whereas SAND varies broadly across the whole series \sylli{and raises a false positive between the first and the second anomaly}. AUC-PR ranks SAND first ($0.225$ against $0.198$), and VUS@20 places the two detectors close together ($0.689$ against $0.608$). The fixed buffer schedule of VUS can thus produce over-optimistic scores that make strong and weak detectors hard to tell apart.}

In this work, we tackle both limitations. We first \textbf{address L1} by introducing \textbf{FastVUS}, which computes VUS at a fraction of the cost (on average $76\times$ faster). Our contribution is orthogonal to parallelizing VUS across series or detectors. Instead, we redesign its per-series computation from scratch. 
FastVUS redesigns \sylli{the VUS computation} by (i) injecting all buffers in a single vectorized pass, (ii) exploiting the monotonicity of the existence reward to prune redundant computations, (iii) casting the confusion matrix over all buffers and thresholds as a single reduced matrix multiplication step, (iv) \sylli{computing VUS for every buffer length} in one pass, and (v) discarding non-anomalous regions with a global mask. We further introduce \textbf{FastVUS-GPU}, which redesigns every step of VUS as batched tensor operations (providing an average speedup of $304\times$ with a maximum error of $5\times10^{-3}$). It replaces sequential \sylli{with parallel existence computation} and chunks long series, keeping evaluation on-device.

\sylli{Second, we \textbf{address L2} by introducing \textbf{AutoVUS}, which builds on VUS by applying \textit{adaptive buffering locally to each anomaly and detector}. FastVUS makes this possible, as it computes VUS for all buffers $0,\ldots,L$ in a single pass.}
This selection requires evaluating how the reward changes across buffer lengths around each anomaly, which is infeasible with the original VUS runtime because it would multiply an already expensive evaluation by the number of anomalies. 
In practice, AutoVUS selects, for each anomaly and detector, the buffer that maximizes gain beyond the analytical random baseline, following the same chance-correction principle as adjusted clustering indices~\cite{romano2016adjusting}. This random-normalized rule limits buffer growth when a larger buffered positive region mainly explains the improvement, while preserving tolerance when the detector genuinely fires near the anomaly. \sylli{For example, in Figure~\ref{fig:intro_adaptive}, AutoVUS separates LOF and SAND clearly ($0.605$ against $0.261$) and selects a different buffer for each detection, i.e., $6$ for the first detection of LOF and $11$ for the second.} AutoVUS thus turns VUS into an event-level diagnostic that reveals the temporal tolerance each \sylli{detection} needs.


Overall, our contributions are as follows:
\begin{compactenum}[$\bullet$]
    \item We propose \textbf{FastVUS}, an exact reformulation of VUS that redesigns \sylli{its entire computation pipeline} \textit{(Section~\ref{sec:autovus})}. 
    \item We introduce \textbf{FastVUS-GPU}, which recasts the pipeline as batched tensor operations while keeping memory complexity low to enable on-device evaluation \textit{(Section~\ref{sec:autovus-gpu})}. 
    \item We propose \textbf{AutoVUS}, which applies \textit{adaptive buffering}, a buffer selection normalized against a random baseline, turning VUS into an event-level measure that reveals the temporal tolerance each detection needs \textit{(Section~\ref{sec:adaptive-buffers})}.
    \item We conduct a \textbf{large-scale evaluation} on the TSB-AD benchmark and synthetic data demonstrating that (i) FastVUS is on average $76\times$ faster than the original VUS on CPU while reproducing it exactly, (ii) FastVUS-GPU reaches an average speedup of $304\times$, approaching the runtime of AUC despite integrating over thresholds and buffer lengths (Figure~\ref{fig:intro_runtime}), \sylli{with} a single approximation \sylli{of} maximum error $5\times10^{-3}$. We further validate FastVUS-GPU's practical impact through a deep-learning pipeline experiment \textit{(Section~\ref{sec:experimental_evaluation})}.
    \item We provide \textbf{case studies} \sylli{and a benchmark-wide comparison} showing that AutoVUS reduces buffer-induced score inflation while preserving meaningful temporal tolerance \textit{(Section~\ref{sec:case})}.
\end{compactenum}

Finally, we summarize our findings \textit{(Section~\ref{sec:conclusion})}. We release our code as open source~\cite{ffvusRepo}.

\section{Background and Related Work}
\label{sec:background}

We first introduce the notation used throughout the paper and review the TSAD evaluation measures (Section~\ref{sec:tsad-measures}). We then explain how to compute VUS and its two limitations (Section~\ref{sec:vus-overview}). Finally, we review the random-normalized measures (Section~\ref{sec:random-measures}) on which AutoVUS builds.


\noindent\textbf{Time Series:} A time series $T \in \mathbb{R}^n$ is a sequence $[T_1,T_2,...,T_n]$ of length $n=|T|$.
A dataset $\mathcal{D}$ is a set of time series with size $|\mathcal{D}|$.

\noindent\textbf{Anomaly Score Sequence:} For a time series $T \in \mathbb{R}^n $, an anomaly detection method $D$ returns an anomaly score sequence $D(T) = s$. The anomaly score has the same length as the time series.

\noindent\textbf{Anomaly Detection Accuracy:} For $T \in \mathbb{R}^n$, the label $y \in \{0,1\}^n$ indicates if the points in $T$ are normal (0) or abnormal (1). We define an accuracy function $Acc: \mathbb{R}^n \times \{0,1\}^n \rightarrow [0,1]$, for which the accuracy score $Acc(D(T), y)$ indicates how accurate $D$ is. 

\noindent\textbf{Anomaly Range:} An anomaly range is a contiguous interval $R_k = [t_k^s, t_k^e)$ where the label is 1, with $t_k^s$ and $t_k^e$ its start and end indices. We denote the ground-truth anomaly ranges by $R = \{R_1, \ldots, R_{N_\alpha}\}$, where $N_\alpha$ is the number of anomalies.

\subsection{TSAD Evaluation Measures}
\label{sec:tsad-measures}
TSAD evaluation measures are either threshold-dependent or threshold-independent. Threshold-dependent measures (e.g., F-score) evaluate binary predictions obtained from a chosen threshold, whereas threshold-independent measures (e.g., AUC and VUS) aggregate performance \sylli{over all} thresholds. \sylli{Both types are either point-based or range-based, that is, they evaluate individual points or anomaly ranges, which gives the four groups we review below.}

\subsubsection{Threshold-Dependent}
\label{sec:threshold-dependent-measures}

\noindent\textbf{\\Point-based.}
\sylli{Threshold-based evaluation converts the anomaly score $s$ into a binary prediction $\hat y(\theta) \in \{0,1\}^n$, with $\hat y_t(\theta) = 1$ if $s_t \geq \theta$ and $0$ otherwise, for a threshold $\theta$ typically set to $\mu(s) + 3\sigma(s)$~\cite{barnett1994outliers}, where $\mu(s)$ is the mean and $\sigma(s)$ the standard deviation. Comparing $\hat y(\theta)$ to the ground-truth labels $y \in \{0,1\}^n$ gives the \textit{True Positives} $\textit{TP} = y^\top \cdot \hat y(\theta)$, \textit{True Negatives} $\textit{TN} = (\mathbf{1}-y)^\top \cdot (\mathbf{1}-\hat y(\theta))$, \textit{False Positives} $\textit{FP} = (\mathbf{1}-y)^\top \cdot \hat y(\theta)$, and \textit{False Negatives} $\textit{FN} = y^\top \cdot (\mathbf{1}-\hat y(\theta))$, where $\mathbf{1}$ is the all-ones vector of length $n$. From these follow the standard \textbf{Precision} $\text{Pr}=\textit{TP}/(\textit{TP}+\textit{FP})$, \textbf{Recall} $\text{Rc}=\textit{TP}/(\textit{TP}+\textit{FN})$, \textbf{False Positive Rate} $\text{FPR}=\textit{FP}/(\textit{FP}+\textit{TN})$, and \textbf{F-score (F)}, the harmonic mean of Precision and Recall.}

\begin{figure*}[!t]
    \centering
    \begin{subfigure}[b]{0.30\textwidth}
        \includegraphics[width=\textwidth, alt={Label before and after the addition of 20 buffers}]{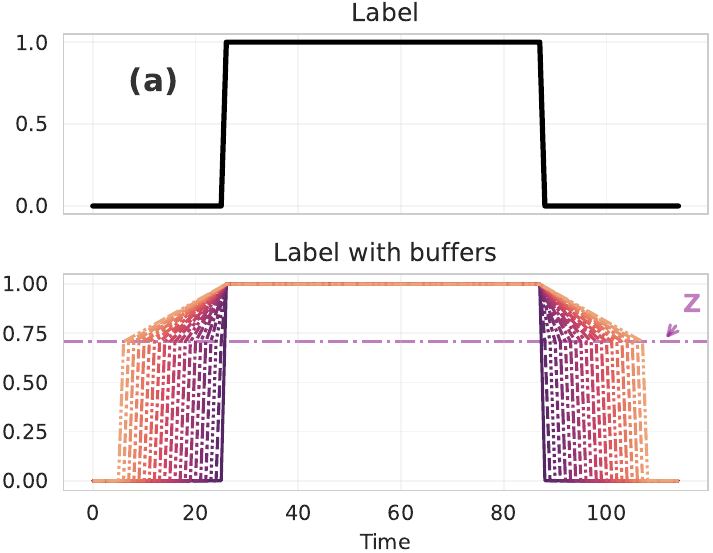}
    \end{subfigure}
    \hfill
    \begin{subfigure}[b]{0.30\textwidth}
        \includegraphics[width=\textwidth, alt={Heatmap showing existence values across buffer and threshold pairs}]{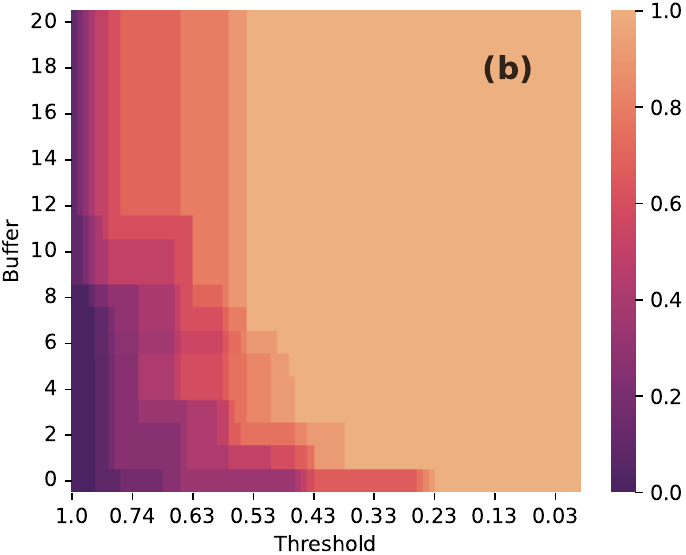}
    \end{subfigure}
    \hfill
    \begin{subfigure}[b]{0.27\textwidth}
        \includegraphics[width=\textwidth, alt={Precision-Recall curves across 20 buffers}]{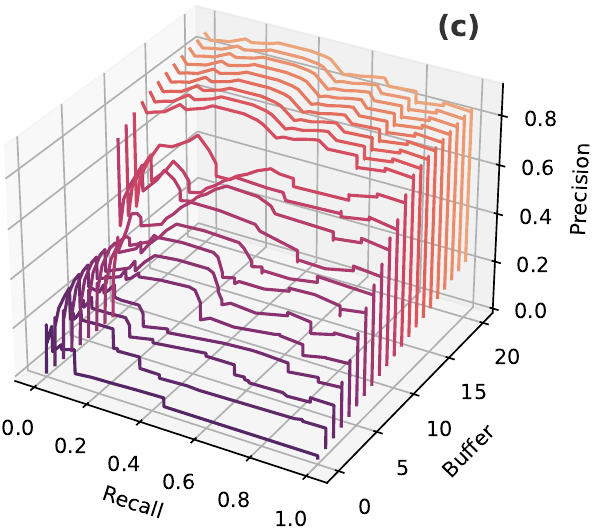}
    \end{subfigure}
    \caption{(a) Label before (top) and after (bottom) adding 20 buffers (buffer height $Z$), (b) existence values over all buffer $\times$ threshold pairs, and (c) resulting Precision-Recall surface across 20 buffers (shown as lines for clarity).}
    \label{fig:triplet_plots}
\end{figure*}

\noindent\textbf{Range-based.}
\sylli{Range-based Precision and Recall~\cite{tatbul_precision_2018} extend the point-based ones to anomaly ranges with four components: an \textit{Existence} reward indicating whether a range was detected (defined in Section~\ref{sec:threshold-independent-measures}), the overlap size, the position of the detection within the range, and the fragmentation of the detection over several predictions, each weighted by application-specific functions. Their harmonic mean is the \textbf{Range F-score (RF)}, which we use as a baseline in Section~\ref{sec:case}. While this formulation accounts for the time-series structure, it still requires parameter tuning and, most importantly, a threshold.}

\sylli{Finally, \textbf{Affiliation}~\cite{huet2022local} affiliates each prediction to its nearest ground-truth anomaly and computes Precision and Recall from the temporal distance between the two, so that near misses still receive credit. Their harmonic mean gives the \textbf{Affiliation F-score (Aff-F)}, our second threshold-dependent baseline in Section~\ref{sec:case}.}

\subsubsection{Threshold-Independent}
\label{sec:threshold-independent-measures}

\noindent\textbf{\\Point-based.} In contrast to the measures above\sylli{, which all require a threshold}, threshold-independent measures evaluate detector performance \sylli{over all} thresholds. The \textbf{Area Under the Receiver Operating Characteristic curve (AUC-ROC)}~\cite{fawcett2006introduction} measures the trade-off between True Positive Rate (TPR) and False Positive Rate (FPR). The True Positive Rate is defined as $\textrm{TPR}(\theta) = \frac{\textit{TP}(\theta)}{\textit{TP}(\theta) + \textit{FN}(\theta)}$, where $\textit{TP}(\theta)$ and $\textit{FN}(\theta)$ are computed using threshold $\theta$ to convert scores to binary predictions. Similarly, the False Positive Rate is $\textrm{FPR}(\theta) = \frac{\textit{FP}(\theta)}{\textit{FP}(\theta) + \textit{TN}(\theta)}$. Given an ordered set of $N_\theta$ thresholds $\Theta = [\theta_1, \theta_2, ..., \theta_{N_\theta}]$, where $0 = \theta_1 < \theta_2 < ... < \theta_{N_\theta} = 1$, AUC-ROC integrates TPR against FPR across these thresholds using the trapezoidal rule. \sylli{Similarly,} the \textbf{Area Under the Precision-Recall curve (AUC-PR)}~\cite{10.1145/1143844.1143874} evaluates the trade-off between Precision and Recall across varying thresholds, integrating $\text{Pr}(\theta)$ against $\text{Rc}(\theta)$ over the same ordered thresholds. \sylli{We use AUC-PR (referred to as \textbf{AUC}) as one of our baselines.}

\noindent\textbf{Range-based.} To address the \sylli{shortcomings of these} point-based measures, the authors in~\cite{paparrizos2022volume} define \textbf{Range-AUC-ROC} and \textbf{Range-AUC-PR} which incorporate subsequence-level evaluation with tolerance for labeling imprecision. \sylli{They} add a buffer region at anomaly boundaries to create a transition zone between normal and abnormal subsequences, rewarding high anomaly scores near outlier boundaries.

A buffer is a tolerance region of length $b$ added before and after each anomaly boundary. Rather than treating labels as binary (0 or 1), buffers extend the labels with continuous values that decay from 1 at the anomaly boundary to a floor $Z = 1/\sqrt{2}$ at distance $b$~\cite{paparrizos2022volume}. This construction allows detectors to receive partial credit for predictions close to, but not perfectly aligned with, the anomaly boundaries. Formally, for an anomaly spanning indices $R_k = [t_k^s, t_k^e)$:
\begin{equation}
\Small
    \tilde y_b(t) =
    \begin{cases}
        \sqrt{1 - \dfrac{|t_k^s-t|}{2b}}, & \text{if } t_k^s - b \leq t < t_k^s \text{ and } \hat y_t(\theta) = 1\\[2pt]
        \sqrt{1 - \dfrac{|t_k^e-t|}{2b}}, & \text{if } t_k^e \leq t < t_k^e + b \text{ and } \hat y_t(\theta) = 1\\[2pt]
        y_t, & \text{otherwise}
    \end{cases}
    \label{eq:vus-buffer}
\end{equation}
Using continuous labels, we can compute $\textit{TP}_b = \tilde y_b^\top \cdot
\hat y(\theta)$, $\textit{FP}_b = (\mathbf{1} - \tilde y_b)^\top \cdot
\hat y(\theta)$, and analogously for $\textit{TN}$ and $\textit{FN}$.

The Existence reward \sylli{introduced for the range-based measures above indicates} whether an anomaly range was detected at least once. Formally, for an anomaly range $R_k$ and a set of predicted ranges $P$, $\text{Existence}(R_k, P) = 1$ if \sylli{some} $P_j \in P$ \sylli{satisfies} $R_k \cap P_j \neq \emptyset$, and 0 otherwise. Range-based TPR incorporates \sylli{it} as follows:
\begin{equation} \label{eq:recall_existence}
\Small
  \textit{TPR}_b = \frac{\textit{TP}_b}{\sylli{P_b}} \cdot \frac{\sum_{R_k \in R} \text{Existence}(R_k, P)}{N_\alpha}
\end{equation}
$R$ and $P$ are sets of ground-truth and predicted anomaly ranges, and \sylli{$P_b$ is the number of positives, defined in~\cite{paparrizos2022volume} as the mean of the original and the buffered positive-label mass}. Range-AUC-ROC and Range-AUC-PR \sylli{follow} the standard AUC formulations, replacing TPR and FPR (or Precision and Recall) with their buffered counterparts. We use Range-AUC-PR (referred to as \textbf{Range-AUC}) as one of our baselines. \sylli{Although these measures are both threshold-independent and time-aware, they add a parameter \sylli{to tune}, the buffer length $b$.}

\subsection{VUS: Computation Overview}
\label{sec:vus-overview}

\sylli{To remove the choice of a single buffer length,} VUS extends Range-AUC by computing the ROC or PR curve for \sylli{every buffer length in a grid. The two-dimensional Range-AUC curve (Precision against Recall, or TPR against FPR) thus becomes a surface in three dimensions, with the buffer length as the third axis}. \sylli{The resulting measure is} the Volume Under the Surface (VUS), for either the ROC (VUS-ROC) or the PR (VUS-PR) surface. In our study, we use VUS-PR, which we refer to as \textbf{VUS} for brevity.

To provide intuition, we walk through its computation steps using Figure~\ref{fig:triplet_plots} as a guide (pseudocode is given in Algorithm~\ref{alg:ff_vus_compute}, Section~\ref{sec:autovus}). First, the original binary labels are extended by adding buffers before and after each anomaly boundary, as shown in Figure~\ref{fig:triplet_plots}a \sylli{(Algorithm~\ref{alg:ff_vus_compute}, line 12)}. Importantly, when buffers are added, regions that would introduce additional false negatives are ignored. Second, for every combination of buffer length and detection threshold, we compute the existence reward shown as a heatmap in Figure~\ref{fig:triplet_plots}b \sylli{(line 13)}. Third, we compute the confusion matrix (TP, FP, TN, FN) for each buffer $\times$ threshold pair \sylli{(line 15)}. Fourth, from these confusion matrices, we derive precision and recall values across all pairs\sylli{, and incorporate the existence reward into recall to ensure detectors are credited (or not) for detecting a larger proportion of anomalies. These values form} a three-dimensional surface over the buffer and threshold dimensions, depicted as separate curves in Figure~\ref{fig:triplet_plots}c \sylli{(line 19)}. \sylli{Fifth}, the volume under this Precision-Recall surface is computed and normalized by the number of buffers, yielding the single VUS score \sylli{(lines 21 and 23)}. The same procedure applies to VUS-ROC, substituting FPR for Precision.

Although VUS is increasingly adopted for TSAD evaluation, it \sylli{still} suffers from \sylli{two} limitations:

\noindent\textbf{(L1) Time-awareness vs. Scalability.} \sylli{By integrating over both thresholds and buffer lengths, VUS introduces a computational bottleneck. Its time complexity is $O(N_\theta N_b n)$~\cite{boniol2025vus}, where $N_\theta$ is the number of thresholds, $N_b = |\mathcal{L}| = L+1$ the number of buffer lengths in the grid $\mathcal{L} = \{0, \ldots, L\}$ up to a maximum $L$, and $n$ the length of the series. On a million-point series, VUS takes roughly 10 seconds against 0.1 seconds for AUC (Section~\ref{sec:scalability_synthetic}), and evaluating the 32 detectors of the TSB-AD benchmark~\cite{liu2024elephant} on its 870 series takes 7 hours and 38 minutes (Section~\ref{sec:acc_speed_tsb}), which makes VUS impractical for studies that evaluate many detectors over many series, such as~\cite{sylligardos2023choose}. The problem compounds when temporal tolerance is examined per anomaly, as anomalies and detections of the same series may need different buffers, since the computation must then be repeated at a finer granularity.}

\newpart{\noindent\textbf{(L2) Fixed buffer limitation.} 
\sylli{Once the buffer length $L$ has been set, VUS} uses the same buffer grid $\mathcal{L}$ for every anomaly and detector. \sylli{Tuning this parameter per series or per anomaly is not practical.} A short anomaly may not require the same tolerance as a long anomaly\sylli{, and a precisely localized detection should not be treated in the same way as a broad one that aligns with the anomaly only under a large buffer (Section~\ref{ssec:benchmark-case})}. 
Moreover, expanding the buffer increases the softened positive region, so the resulting increase in score may partly \sylli{correspond to} a higher chance-level reward rather than better temporal localization. \sylli{For example, a uniformly random score obtains on average over TSB-AD a VUS of $0.03$ at $L=0$ and of $0.16$ at $L=128$, where AutoVUS at $L=128$ gives it $0.04$. This inflation makes detector performances harder to tell apart.} These limitations motivate selecting a buffer separately for each anomaly-detector combination\sylli{, a local search that requires an efficient computation to be practical}. A chance-corrected criterion is needed to identify the buffer whose observed gain exceeds the gain expected from a random detector.}

\subsection{Random-Normalized Measures}
\label{sec:random-measures}

\newpart{As mentioned above, \sylli{expanding the buffer enlarges the softened positive region, so the VUS score can rise through a higher chance-level reward rather than through better localization}. This issue is well studied in clustering, where even a random partition may accidentally match part of the ground truth, producing nonzero scores for measures such as the Rand index~\cite{rand1971objective} and mutual information~\cite{vinh2010information}. Adjusted measures \sylli{subtract} the expected score under a random model:
\begin{equation*}
    \mathrm{Adj}(V)
    =
    \frac{V - \mathbb{E}[V_{\mathrm{rand}}]}
    {V_{\max} - \mathbb{E}[V_{\mathrm{rand}}]},
\end{equation*}
where $V$ is the value of the measure, $V_{\mathrm{rand}}$ its value under the random model, and $V_{\max}$ its value under perfect agreement.
The adjusted Rand index (ARI) applies this correction to pair-counting agreement~\cite{hubert1985comparing}, and adjusted mutual information (AMI) applies it to information-theoretic agreement~\cite{vinh2010information}. \sylli{Such} adjusted-for-chance clustering comparison measures have been systematically studied~\cite{romano2016adjusting}. We use the same principle \sylli{in Section~\ref{sec:adaptive-score}} to normalize buffer-induced gains against the score expected from a random detector.}

\section{Proposed Approach: \textit{A Fast and Adaptive VUS}}
\label{sec:methodology}

\sylli{In this section, we introduce our two contributions, one per limitation of VUS, i.e., its computational cost (L1) and its fixed buffer (L2). Our first contribution is} \textbf{FastVUS}, an algorithm that computes the existing VUS measure (Section~\ref{sec:vus-overview}) returning exactly the same scores \sylli{without} its computational bottlenecks, \textbf{addressing L1}\sylli{. Its GPU variant,} \textbf{FastVUS-GPU}\sylli{, runs the same computation fully in parallel on-device with one approximation in the existence term. Our second contribution is} \textbf{AutoVUS}, \sylli{an adaptive extension of VUS}, made practical by FastVUS. Instead of \sylli{one} buffer length \sylli{for} the entire series, AutoVUS \sylli{selects one} for each anomaly and detector, making VUS an interpretable, event-level diagnostic, \textbf{addressing L2}.

\subsection{FastVUS: An Exact Reformulation of VUS}
\label{sec:autovus}

FastVUS \sylli{redesigns each of the five steps of VUS (Section~\ref{sec:vus-overview}) and adds one optimization on top, as} summarized in Algorithm~\ref{alg:ff_vus_compute}. Every step is exact.
\begin{enumerate}
    \item \textbf{Buffer injection}: inject all buffers up to the maximum length in a single pass.
    \item \textbf{Existence computation}: fraction of anomalies detected, skipping redundant buffer $\times$ threshold evaluations.
    \item \textbf{Confusion matrix}: formulation \sylli{that yields the TP, FP, FN, TN} across all buffer $\times$ threshold pairs \sylli{in one step}.
    \item \sylli{\textbf{Precision and recall}: the precision-recall surface over all buffer $\times$ threshold pairs.}
    \item \sylli{\textbf{VUS integration}: integrate the surface into VUS, yielding the full VUS$(b)$ curve at no extra cost.}
\end{enumerate}
\sylli{The optimization (Section~\ref{sec:global-mask}) is the \textbf{global mask}, which drops non-anomalous regions.}

\begin{algorithm}[t]
\footnotesize
\caption{FastVUS Main Computation Loop}
\label{alg:ff_vus_compute}
\begin{algorithmic}[1]
\REQUIRE Label $y \in \{0,1\}^n$, Anomaly Score $s \in \mathbb{R}^n$
\STATE $R \leftarrow \text{GetAnomalyRanges}(y), \quad \Theta \leftarrow \text{Unique}(s)$
\STATE \textcolor{blue}{\# Optional (see Section~\ref{sec:global-mask})}
\IF{GlobalMask is Active}
    \STATE \textcolor{blue}{\# Count FPs in non-buffer regions before discarding them}
    \STATE $M_{\textrm{safe}} \leftarrow \text{CreateSafeMask}(y, R)$
    \STATE $\textit{FP}_{\textrm{ext}} \leftarrow \text{CountExceeding}(s[\neg M_{\textrm{safe}}], \Theta)$
    \STATE \textcolor{blue}{\# Filter input to keep only buffer regions}
    \STATE $y \leftarrow y[M_{\textrm{safe}}], \quad s \leftarrow s[M_{\textrm{safe}}]$
    \STATE \textcolor{blue}{\# Re-calculate ranges on reduced vector}
    \STATE $R \leftarrow \text{GetAnomalyRanges}(y)$
\ENDIF
\STATE $Y \leftarrow \text{AddBuffers}(y, R)$
\STATE $E \leftarrow \text{ComputeExistence}(Y, s, \Theta)$
\STATE $S \leftarrow (s \geq \Theta)$
\STATE $\textit{TP}, \textit{FP}, \textit{FN} \leftarrow \text{ConfusionMatrix}(Y, S)$
\IF{GlobalMask is Active}
    \STATE $\textit{FP} \leftarrow \textit{FP} + \textit{FP}_{\textrm{ext}}$ \COMMENT{Add back FP from filtered regions}
\ENDIF
\STATE $\text{Pr}, \text{Rc} \leftarrow \text{PrecisionRecallCurve}(\textit{TP}, \textit{FP}, \textit{FN}, E)$
\STATE \textcolor{blue}{\# Per-buffer AUC: one area per buffer $b \in \mathcal{L}$}
\STATE $A \leftarrow \text{AUC}(\text{Rc}, \text{Pr})$
\STATE \textcolor{blue}{\# Cumulative mean: $\text{VUS}(b)$ is the mean of areas up to buffer $b$}
\STATE $\text{VUS} \leftarrow \text{CumulativeMean}(A)$
\STATE \textcolor{blue}{\# Full curve; $\text{VUS}(L)$ is the standard scalar}
\RETURN $\text{VUS}$
\end{algorithmic}
\end{algorithm}

\subsubsection{Buffer Injection}
\label{sec:buffer-injection}

Buffers are the core of VUS. Around every anomaly, VUS adds a soft tolerance region so that a detection close to a boundary is rewarded. The original implementation inserts these buffers one anomaly at a time and rewrites the label once per buffer length, which becomes a bottleneck. We instead compute all buffers at once through a closed-form function, shared by FastVUS and FastVUS-GPU.

\begin{figure}
    \centering
    \includegraphics[width=\linewidth]{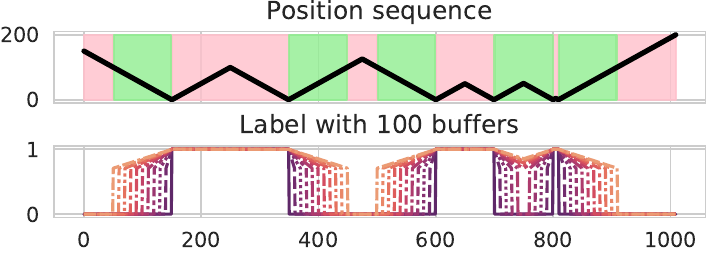}
    \Description{Proposed vectorized buffer injection.}
    \vspace{-0.5cm}
    \caption{Proposed vectorized buffer injection. Top: the \textit{position} sequence (distance of each point to its closest anomaly boundary), with buffered regions in green and unaffected regions in red. Bottom: the resulting buffered label.}
    \label{fig:buffer_injection}
\end{figure}

We first compute a \textit{position} sequence from the binary label (top of Figure~\ref{fig:buffer_injection}). For every point outside an anomaly, position gives its distance to the closest anomaly boundary; points inside an anomaly have position $0$ (so they keep label $1$). It is obtained in parallel across all time points. Given a buffer length $b$, the buffered label follows
\begin{equation*}
    \Small
    \tilde y_b(t) =
    \begin{cases}
        \sylli{\sqrt{1 - (1 - Z^2)\, \dfrac{\text{pos}(t)}{b}}}, & \text{if } \text{pos}(t) \leq b,\\
        y_t, & \text{otherwise,}
    \end{cases}
\end{equation*}
where $Z$ controls how rewarding the buffer is\sylli{. The reward} starts at $1$ on the boundary and decays to $Z$ at distance $b$ \sylli{(the profile of Equation~\ref{eq:vus-buffer})}, while points farther than $b$ from any anomaly keep their original label. Where two buffers overlap, a point takes the maximum of the two values. To obtain every buffer at once, we evaluate this function for all lengths $b \in \mathcal{L}$, producing the full buffered-label matrix $Y$ in a single vectorized pass \sylli{(bottom of Figure~\ref{fig:buffer_injection})}.

\subsubsection{Existence Computation}
\label{sec:existence}

The \textit{existence} reward is the fraction of anomalies detected by at least one point of the score\sylli{, and it was} the main bottleneck of VUS. The standard definition iterates over every anomaly for every buffer and threshold, at a cost of $O(N_\theta N_b N_\alpha)$. However, we observe that \textit{existence} is \textit{monotonic} in both dimensions. A detection at a given threshold persists at every lower threshold, and a detection at a given buffer length persists at every larger buffer. We exploit \sylli{it} as follows:

\noindent\textbf{Threshold propagation.} For each anomaly we take the maximum score inside its range, its \textit{detection threshold}, and mark it as detected for that threshold and all lower ones at once. This removes the threshold loop, reducing the cost from $O(N_\theta N_b N_\alpha)$ to $O(N_b N_\alpha)$.

\noindent\textbf{Buffer skipping.} We locate where the maximum score falls within the buffered region. If it falls inside the original anomaly \sylli{(pink in Figure~\ref{fig:conf_matrix_mask})}, the anomaly is detected for \textit{every} buffer length, so we fill all buffers at once and move to the next anomaly. If it falls inside a buffer \sylli{(purple in Figure~\ref{fig:conf_matrix_mask})}, we compute the exact buffer index at which detection begins, mark every larger buffer as detected, and recurse on the smaller buffers. Processing buffers from largest to smallest, each step either terminates or shrinks the remaining range, so most buffer evaluations are never performed. \sylli{Growing buffers may cause neighboring anomalies to merge.} We track these merges and adjust the anomaly count per buffer so the fractions stay exact.
\newpart{On TSB-AD, this procedure skips on average $93.9\%$ of the $N_b N_\alpha$ evaluations, while computing the exact existence reward.}

\subsubsection{Confusion Matrix}
\label{sec:conf-matrix}

For every buffer and threshold pair, VUS builds a confusion matrix ($\textit{TP}$, $\textit{FP}$, $\textit{FN}$, $\textit{TN}$). The original VUS fills it with a double loop over the $N_b$ buffers and the $N_\theta$ thresholds. Our key contribution is to recast the entire computation as a single matrix multiplication $\textit{TP} = Y S^\top$ between the buffered-label matrix $Y$ and the thresholded-score matrix $S$\sylli{, replacing $N_b N_\theta$ separate computations by one}. $\textit{FP}$, $\textit{FN}$, and $\textit{TN}$ then follow from element-wise operations. To the best of our knowledge, this is the first time VUS is expressed this way, and \sylli{this reformulation} lets the computation \sylli{use} optimized dense linear algebra on both CPU and GPU.

\begin{figure}
    \centering
    \includegraphics[width=\linewidth]{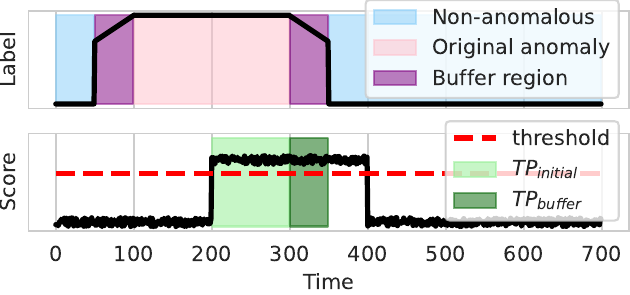}
    \Description{Confusion matrix computation.}
    \vspace{-0.5cm}
    \caption{Confusion matrix computation. Top: the label splits into three regions. Bottom: the regions \sylli{used to compute} $\textit{TP}$.}
    \label{fig:conf_matrix_mask}
\end{figure}

We further exploit the structure of the buffered labels to make this multiplication cheap. Each label splits into three regions (Figure~\ref{fig:conf_matrix_mask}): \sylli{(1)} the non-anomalous region ($\tilde y_b(t) = 0$), \sylli{(2)} the original anomaly ($\tilde y_b(t) = 1$), and \sylli{(3)} the buffer ($0 < \tilde y_b(t) < 1$). Non-anomalous regions contribute nothing to $\textit{TP}$ and can be excluded, an optimization introduced in the journal version of VUS~\cite{boniol2025vus}. \sylli{The original anomaly, in turn, yields} $\tilde y_b(t) = 1$ for every buffer, so \sylli{its} contribution is constant across buffers and reduces to a single dot product. \sylli{Together, these observations give the} final form of $\textit{TP}$, a constant term $\textit{TP}_\text{initial}$ over \sylli{the original anomalies} plus a matrix multiplication $\textit{TP}_\text{buffer}$ restricted to the buffer regions of total length $2 L N_\alpha$:
\begin{align*}
    \textit{TP} &= \textit{TP}_\text{initial} + \textit{TP}_\text{buffer} \\
    &= y[M_a] \cdot S[:, M_a]^\top + Y[:, M_b] \cdot S[:, M_b]^\top,
\end{align*}
where $y$ is the original binary label, $M_a$ the anomaly mask, and $M_b$ the buffer mask. The total cost is $O(N_b L N_\theta N_\alpha) + O(n(N_b + N_\theta))$. Since anomalies typically occupy a small fraction of the series ($2 L N_\alpha \ll n$), this is far below the $O(N_\theta N_b n)$ of \sylli{the original VUS}.
One might expect that a general-purpose sparse library could \sylli{replace these computations}, since both matrices are sparse. \sylli{Yet} our time-series-specific computation is about $9\times$ faster \sylli{(median per-series ratio on TSB-AD)} than sparse libraries on both CPU (\texttt{scipy.sparse}) and GPU (\texttt{cuSPARSE})\sylli{. In any case,} the confusion matrix is one of several steps of VUS, so replacing it alone would not remove the bottleneck\sylli{, as the ablation study in Section~\ref{sec:acc_speed_tsb} shows}.

\subsubsection{\sylli{Precision and Recall}}
\label{sec:pr-vus}

With the confusion matrix in hand, \sylli{precision} follows from $\textit{TP}$ and $\textit{FP}$, and recall from $\textit{TP}$ and $\textit{FN}$, scaled by the existence reward. \sylli{Across} the $N_b$ buffers and the $N_\theta$ thresholds, they form the precision-recall surface\sylli{.}

\subsubsection{\sylli{VUS Integration}}
\label{sec:vus-integration}

\sylli{VUS-PR is defined as} the volume under this surface. For each buffer, we take the area under its precision-recall curve, and average these areas over all buffers to obtain the final score. Its ROC variant, VUS-ROC, requires a single change at this final step, replacing Precision with the False Positive Rate. Every earlier step remains identical \sylli{and FastVUS implements} both.

\newpart{This averaging \sylli{over buffers} hides a structure we exploit. \sylli{We retain the area of every buffer and divide the running sum of areas by the number of buffers so far, producing the curve $\textit{VUS}(b)$ for every buffer length $b$ from $0$ to $L$ in a single pass. This is possible because buffer lengths are nested, and a larger buffer includes every smaller one, so computing VUS at the maximum length $L$ evaluates the precision-recall curve of every shorter buffer, which the original implementation of VUS~\cite{paparrizos2022volume} discards.} The standard VUS scalar is recovered as the last element\sylli{,} $\textit{VUS}(L)$\sylli{.}
VUS thus becomes a function of the buffer length rather than a single value. This curve is what makes the per-anomaly buffer selection of Section~\ref{sec:adaptive-buffers} possible. Without it, recovering VUS at each candidate buffer length would require recomputing the measure for every length.}

\subsubsection{Global Mask}
\label{sec:global-mask}

Anomalies are rare, so most of a time series contributes only true negatives and false positives\sylli{, which we exploit with the \textit{global mask}}. We remove every non-anomalous region from the label, keeping only the anomalies and their buffers, and run the entire pipeline on this shorter sequence. Exactness is preserved because the removed regions can only produce false positives. We count these once, before masking, and add them back at the end. Across TSB-AD, the kept region (\sylli{the original anomalies} together with their buffers) spans on average $24\%$ of the series, so the global mask removes the remaining $76\%$ \sylli{(its contribution to the speedup is measured in the ablation study of Section~\ref{sec:acc_speed_tsb})}.

\subsection{FastVUS-GPU: A Fully Parallel Formulation}
\label{sec:autovus-gpu}
\sylli{We now introduce FastVUS-GPU, a fully vectorized reformulation of FastVUS} that executes the entire computation on the GPU as batched tensor operations\sylli{,} keeping the evaluation on-device. \sylli{Two parts are GPU-specific, the existence computation} and a memory-chunking step. All other steps are identical to FastVUS (Section~\ref{sec:autovus}) and remain exact.

\subsubsection{Vectorized Existence}
\label{sec:gpu-existence}

A GPU applies a single operation to an entire batch at once rather than executing steps in sequence. The existence computation of FastVUS, however, saves work by proceeding sequentially, skipping buffers based on intermediate results, which parallel execution cannot express. We therefore recast existence as a fully parallel computation \sylli{to take full advantage of the GPU}.

We first build a staircase-like sequence from the buffered label by assigning a unique increasing index to each anomaly segment\sylli{. In practice, this is} a copy of the label in which each anomaly carries its own index ($1, 2, 3, \ldots$) rather than a $1$\sylli{. Normal} regions stay $0$. We build one such sequence per buffer \sylli{with a cumulative sum}, \sylli{which gives} an $N_b \times n$ tensor. \sylli{With this staircase tensor, we extract the maximum score of each anomaly in a single grouped reduction, since every point is tagged with the segment it belongs to.} This yields an $N_b \times N_\alpha$ tensor holding one per-segment maximum for every anomaly and buffer. An anomaly is detected at a given threshold if and only if its maximum exceeds \sylli{that threshold}, so the existence reward is the fraction of segments whose maximum clears the threshold, an $N_b \times N_\theta$ tensor. This is the same criterion as the detection threshold of the CPU version (Section~\ref{sec:existence}), now evaluated in parallel.

\sylli{The vectorized existence computation} is the single source of approximation in FastVUS-GPU\sylli{, apart from single-precision rounding on the GPU}. \sylli{When} two buffered anomaly segments are non-overlapping yet adjacent, they receive the same index and merge into one, whereas the sequential version keeps them distinct. \sylli{Empirically, the} error stays below $5\times10^{-3}$ \sylli{on TSB-AD at $L=128$} (Section~\ref{sec:acc_speed_tsb}).

\subsubsection{Memory Chunking}
\label{sec:gpu-chunking}

\newpart{Processing every buffer and threshold in parallel materializes large tensors, namely the $N_b \times n$ buffered labels and the $N_\theta \times n$ score mask, which can exceed the available VRAM on long series. \sylli{To tackle this, we} split the series into chunks, cutting only within the non-anomalous gaps so that each anomaly and its buffers remain within a single chunk. Each chunk contributes its confusion-matrix counts and its detected and total anomaly counts, which we combine at the end. Because the cuts fall outside every buffer, the combined result is identical to processing the whole series at once\sylli{, that is,} chunking affects memory, not the score.}

\subsection{AutoVUS: An Adaptive, Interpretable Measure}
\label{sec:adaptive-buffers}

\newpart{FastVUS and FastVUS-GPU remove the computational bottleneck of VUS\sylli{, which also lets us address a second limitation of VUS}. Standard VUS averages performance over the same buffer grid for every anomaly and detector, even though the temporal tolerance \sylli{each detection needs} may vary across events. A precisely localized detection may require little or no tolerance, whereas an early or delayed response may align with the anomaly only after a wider buffer is allowed. Identifying this tolerance separately for every anomaly requires evaluating the complete buffer-specific performance curve at the event level, which would be prohibitively expensive with the original VUS. FastVUS makes this analysis practical by returning the full curve in a single pass. Building on this, AutoVUS selects an anomaly-specific buffer using gains normalized against a random baseline and reports a chance-adjusted score that remains interpretable across the resulting buffer configurations.}

\subsubsection{Problem Setup}
\label{sec:adaptive-setup}

\newpart{AutoVUS takes as input a binary label $y\in\{0,1\}^{n}$, a detector score $s\in\mathbb{R}^{n}$, and a maximum buffer length $L$. Let the anomalous intervals in $y$ be $R_1,\ldots,R_{N_\alpha}$, and let $\mathcal{L}=\{0,\ldots,L\}$ be the candidate buffer grid. Standard VUS evaluates every anomaly over this common grid and averages the resulting performance. AutoVUS instead selects a per-anomaly buffer vector $b^*=(b_1^*,\ldots,b_{N_\alpha}^*)$, where $b_k^*\in\mathcal{L}$ is the maximum temporal tolerance selected for anomaly $R_k$ \sylli{given the score $s$, i.e., it selects one buffer per anomaly and detector pair. Thus, unlike VUS, where the maximum buffer length $L$ is the tolerance granted to every anomaly, in AutoVUS $L$ is only an upper bound on the tolerance granted.}

The selected buffers also provide an event-level diagnostic of detector behavior. A small $b_k^*$ indicates that the detector localizes anomaly $R_k$ precisely, whereas a larger value indicates that the detector response aligns with the event only after wider temporal tolerance is allowed.}

\subsubsection{Random-Normalized Buffer Selection}
\label{sec:adaptive-selection}

\newpart{For each anomaly $R_k$, we define a local window $W_k$ by splitting the gaps between $R_k$ and its neighboring anomalies. This isolates the region in which detector responses can be attributed to $R_k$. Within $W_k$, let $C_k(b;s)=(b+1)^{-1}\sum_{b'=0}^{b}A_k(b';s)$ denote the cumulative local VUS curve, where $A_k(b;s)$ is the marginal AUC obtained at buffer length $b$. Since FastVUS returns the cumulative curve, the marginal values are recovered exactly by de-cumulation: $A_k(0;s)=C_k(0;s)$ and $A_k(b;s)=(b+1)C_k(b;s)-bC_k(b-1;s)$ for $b\ge1$.

Let $\tilde y_{k,b}\in[0,1]^{|W_k|}$ denote the local buffered label obtained by applying buffer length $b$ to anomaly $R_k$ within $W_k$. Under a random-ranking model in which detector scores are independent of the label, the expected AUC equals the average positive mass of the corresponding buffered label. We therefore compute the analytical random baseline as
\begin{equation*}
    \Small
g_k(b)=\mathbb{E}\!\left[A_k(b;s_{\mathrm{rand}})\mid\tilde y_{k,b}\right]=\frac{1}{|W_k|}\sum_{t\in W_k}\tilde y_{k,b}(t),
\end{equation*}
where $s_{\mathrm{rand}}$ denotes a random detector score independent of the label. This baseline is computed directly from the buffered label\sylli{, without} Monte Carlo sampling. Since increasing the buffer enlarges the softened positive region, $g_k(b)$ generally increases with $b$. Directly maximizing $A_k(b;s)$ would therefore favor longer buffers even when the apparent improvement arises only from a larger chance-level reward.

\begin{figure}
\centering
\includegraphics[width=\linewidth]{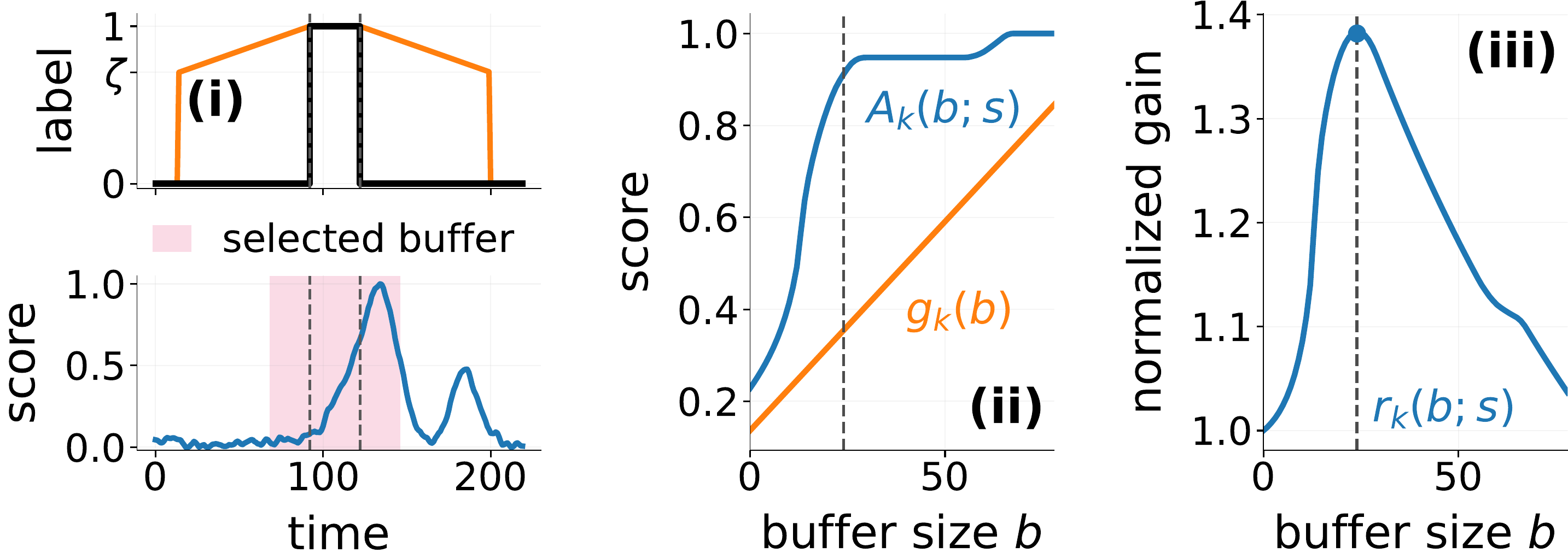}
\Description{Adaptive buffer selection.}
\vspace{-0.3cm}
\caption{\sylli{Adaptive buffer selection. (i) Label $y$, score $s$, and the buffer selected for anomaly $R_k$. (ii) Marginal AUC $A_k(b;s)$ and random baseline $g_k(b)$ over buffer length $b$. (iii) Random-normalized gain $r_k(b;s)$, its smallest maximizer gives $b_k^*$.}}
\label{fig:adaptive_buffer_selection}
\end{figure}

To isolate the gain attributable to the detector response, we compare each buffer with the no-buffer case. Let $\Delta A_k(b;s)=A_k(b;s)-A_k(0;s)$ denote the detector's buffer-induced gain and let $\Delta g_k(b)=g_k(b)-g_k(0)$ denote the corresponding random gain. We select the smallest buffer that maximizes the random-normalized gain:
\begin{equation*}
    \Small
b_k^*=\min\text{ }\operatorname*{arg\,max}_{b\in\mathcal{L}}r_k(b;s),\quad r_k(b;s)=\frac{\Delta A_k(b;s)+1}{\Delta g_k(b)+1}.
\end{equation*}

The additive constant makes the objective well-defined at $b=0$, where $\Delta A_k(0;s)=\Delta g_k(0)=0$ and therefore $r_k(0;s)=1$. Selecting the smallest maximizer resolves ties and favors the least temporal tolerance supported by the detector. In particular, if no positive buffer produces a gain above its random counterpart, then $r_k(b;s)\le1=r_k(0;s)$ for all $b$, and the selected buffer is $b_k^*=0$.

Figure~\ref{fig:adaptive_buffer_selection} illustrates the selection procedure. As the buffer length increases, $A_k(b;s)$ increases because more detection may fall within the buffered label, but $g_k(b)$ also increases as the softened positive region grows. AutoVUS therefore selects the peak of the random-normalized gain curve rather than the largest raw marginal AUC.}

\newpart{The value $1$ is the chance-level reference for buffer-induced gain, since $r_k(b;s)-1=\frac{\Delta A_k(b;s)-\Delta g_k(b)}{\Delta g_k(b)+1}$, so $r_k(b;s)>1$ if and only if the detector obtains more buffer-induced gain than expected from the random baseline. For a random detector independent of the label, $\mathbb{E}[A_k(b;s_{\mathrm{rand}})]=g_k(b)$ and therefore $\mathbb{E}[\Delta A_k(b;s_{\mathrm{rand}})]=\Delta g_k(b)$. Since $\Delta g_k(b)$ is fixed by the label and buffer construction, $\mathbb{E}[r_k(b;s_{\mathrm{rand}})]=1$ for every fixed $b$. The null curve is thus flat in expectation, and longer buffers do not receive a systematic advantage solely because they increase the positive mass of the buffered label.}

\subsubsection{\texorpdfstring{\sylli{Chance-Adjusted VUS}}{Chance-Adjusted VUS}}
\label{sec:adaptive-score}

\newpart{The selection step produces the per-anomaly buffer vector $b^*=(b_1^*,\ldots,b_{N_\alpha}^*)$. Given $b^*$, we perform the VUS buffer sweep over every $b\in\mathcal{L}$, but cap the expansion of each anomaly $R_k$ at its selected buffer, so at sweep level $b$ anomaly $R_k$ uses buffer length $c_k(b)=\min\{b,b_k^*\}$. Let $\tilde y_{b^*}^{(b)}\in[0,1]^n$ denote the resulting global buffered label at sweep level $b$, and let $A_{b^*}(b;s)$ denote the AUC computed from this label and detector score $s$. \sylli{The adapted VUS score then is:}
\begin{equation*}
    \Small
\mathrm{VUS}_{b^*}(s)
=
\frac{1}{N_b}
\sum_{b\in\mathcal{L}}
A_{b^*}(b;s).
\end{equation*}

Different selected buffer vectors produce different sequences of global buffered labels and, therefore, different chance baselines. Under a random-ranking model, the expected AUC at sweep level $b$ equals the average positive mass of the corresponding buffered label, $g_{b^*}(b)=\frac{1}{n}\sum_{t=1}^{n}\tilde y_{b^*}^{(b)}(t)$. The analytical random baseline for the full configuration is the mean over the sweep, $\bar g_{b^*}=\frac{1}{N_b}\sum_{b\in\mathcal{L}}g_{b^*}(b)$. We report the chance-adjusted AutoVUS score:
\begin{equation*}
    \Small
\mathrm{AutoVUS}_{b^*}(s)
=
\frac{
\mathrm{VUS}_{b^*}(s)
-
\bar g_{b^*}
}{
1-\bar g_{b^*}
}.
\end{equation*}
\sylli{This follows the adjustment-for-chance form, that is, observed performance minus expected random performance}, divided by perfect performance minus expected random performance. Both terms \sylli{use} the same anomaly-specific buffer caps.}

\subsubsection{Calibration and Comparability}
\label{sec:adaptive-calibration}

\newpart{We first establish calibration for any fixed per-anomaly buffer vector $b^*$. Since $A_{b^*}(b;s)\le1$ at every sweep level, $\mathrm{VUS}_{b^*}(s)\le1$ and therefore $\mathrm{AutoVUS}_{b^*}(s)\le1$.

For the chance endpoint, suppose $s_{\mathrm{rand}}$ is a random detector independent of the label. At every fixed sweep level, $\mathbb{E}[A_{b^*}(b;s_{\mathrm{rand}})]=g_{b^*}(b)$, so for any fixed selected-buffer vector $\mathbb{E}[\mathrm{VUS}_{b^*}(s_{\mathrm{rand}})\mid b^*]=\bar g_{b^*}$ and therefore $\mathbb{E}[\mathrm{AutoVUS}_{b^*}(s_{\mathrm{rand}})\mid b^*]=0$. Random performance thus maps to $0$ in expectation for every fixed configuration.

For the perfect endpoint, consider a perfectly localized detector, whose anomalous points all rank above the non-anomalous ones, so $A_k(0;s)=1$ for every anomaly $R_k$. For any $b>0$, $A_k(b;s)\le1$ and $g_k(b)\ge g_k(0)$ give $r_k(b;s)\le1=r_k(0;s)$, so the smallest maximizer is $b_k^*=0$ for every anomaly. Then $c_k(b)=0$ at every sweep level, every global buffered label $\tilde y_{b^*}^{(b)}$ equals the original binary label, and the detector obtains $A_{b^*}(b;s)=1$ throughout. Hence $\mathrm{VUS}_{b^*}(s)=1$ and $\mathrm{AutoVUS}_{b^*}(s)=1$.

The chance adjustment also provides comparability across detectors that select different buffer vectors. Suppose two detectors $s_1$ and $s_2$ select $b_1^*$ and $b_2^*$, respectively. Their raw adapted VUS values are evaluated using \sylli{different buffers per anomaly}, and therefore may have different random baselines. AutoVUS removes each selected configuration's expected random performance and normalizes by its remaining gap to perfection. Thus, conditional on each selected-buffer configuration, $0$ represents chance-level performance in expectation and $1$ represents perfect performance.

The preceding calibration is conditional on a fixed selected-buffer vector. Since $b^*$ is selected from the observed detector scores, the maximization step may capture a positive finite-sample fluctuation for a random detector. The unconditional expectation of the complete adaptive procedure, therefore, need not be exactly zero \sylli{($0.04$ on TSB-AD, Section~\ref{sec:vus-overview})}.}

\section{FastVUS Experimental Evaluation}
\label{sec:experimental_evaluation}


\sylli{We evaluate FastVUS along three axes, its equivalence to the original VUS (Section~\ref{sec:acc_speed_tsb}), its runtime (Sections~\ref{sec:acc_speed_tsb} and~\ref{sec:scalability_synthetic}), and its behavior inside a real evaluation workflow (Section~\ref{sec:practicality}).} Our experiments use the TSB-AD benchmark~\cite{liu2024elephant} (\sylli{$870$} univariate series and $32$ detectors)\sylli{, except for the deep-learning pipeline (Section~\ref{sec:practicality}), which uses its predecessor TSB-UAD~\cite{paparrizosTSB2022}}. To study scalability under controlled conditions, we also generate synthetic series varying three characteristics: (a) the series length, (b) the number of anomalies, and (c) the average anomaly length. Each characteristic is varied while the other two are held fixed, and every configuration is averaged over $10$ random datasets.

We compare FastVUS against the original VUS~\cite{paparrizos2022volume}\sylli{, in its optimized implementation~\cite{boniol2025vus}}, and against four other TSAD evaluation measures: AUC, Range-AUC~\cite{paparrizos2022volume}, the Range F-score~\cite{tatbul_precision_2018}, and Affiliation~\cite{huet2022local}. \sylli{In this section we use a buffer length of $L=128$, a computationally challenging but realistic setting. In the case studies of Sections~\ref{ssec:benchmark-case} to~\ref{ssec:case-multi}, which examine the VUS values themselves, we use $L=20$, and in Section~\ref{ssec:benchmark-wide} we vary $L$ from $2$ to $128$.} \sylli{Both VUS and FastVUS use $250$ evenly spread thresholds as in the original VUS.} \newpart{All \sylli{experiments}, including the GPU variant, run on a single laptop (Intel Core Ultra 7 265H, 64\,GB of RAM, and an NVIDIA RTX PRO 1000 Blackwell GPU with 8\,GB of VRAM)\sylli{, except for the synthetic experiments, which} run on a cluster with an AMD EPYC 7352 CPU (256\,GB of RAM) and an NVIDIA H100 NVL GPU (94\,GB of VRAM).}

\begin{figure}[t]
    \centering
    \begin{subfigure}[b]{0.38\linewidth}
        \centering
        \caption{Error to VUS}
        \includegraphics[width=\linewidth]{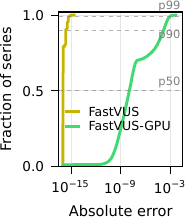}
        \label{fig:error_analysis_tsb}
    \end{subfigure}
    \hfill
    \begin{subfigure}[b]{0.60\linewidth}
        \centering
        \caption{Per-step ablation}
        \includegraphics[width=\linewidth]{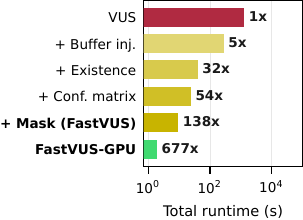}
        \label{fig:ablation}
    \end{subfigure}
    \vspace{-0.7cm}
    \Description{Accuracy and per-step ablation of FastVUS on TSB-AD.}
    \caption{\sylli{Accuracy and ablation on TSB-AD. (a) Absolute error to the original VUS, at most $2\times10^{-15}$ for FastVUS and $5\times10^{-3}$ for FastVUS-GPU. (b) Each technique added to the previous ones, speedup as total-time ratio to the original VUS.}}
    \label{fig:accuracy_ablation}
\end{figure}

\subsection{Accuracy and Speed on TSB-AD}
\label{sec:acc_speed_tsb}

We first compare all evaluation measures on the real-world data of TSB-AD. Figure~\ref{fig:intro_runtime} reports the average runtime per time series and the total time across the benchmark. The original VUS is the slowest, taking $7$ hours and $38$ minutes over the benchmark, which FastVUS reduces to $7$ minutes and $41$ seconds and FastVUS-GPU to $56$ seconds. We report speedups as the average of the per-series ratios, which weights every series equally and thus does not equal the ratio of the totals above. On this basis, FastVUS is $75.91\times$ faster than VUS and FastVUS-GPU $303.75\times$. FastVUS-GPU is faster than the other range-aware measures, Range-AUC ($98.25\times$) and the Range F-score ($94.97\times$), and comes within an order of magnitude of the point-based AUC ($1137\times$) and Affiliation ($684.75\times$).

We measure the error as the absolute difference between our implementations and the original VUS (Figure~\ref{fig:error_analysis_tsb}). FastVUS is numerically exact, with errors at floating-point precision (a maximum of $2\times10^{-15}$), while FastVUS-GPU exhibits a maximum error of $5\times10^{-3}$, due to the approximate existence term (Section~\ref{sec:gpu-existence}). This is the largest error across the $27{,}840$ series and detector pairs of TSB-AD. 

\newpart{Finally, we quantify how much each computation step contributes to the overall speedup. \sylli{In this ablation study we use a single detector (not all detectors as above, so the speedups differ) and add our techniques one at a time, starting from the original VUS. Figure~\ref{fig:ablation} shows that every step of FastVUS contributes, but two stand out.} The optimized existence computation yields the largest jump, a sevenfold reduction (from $\sylli{5}\times$ to $32\times$), confirming that existence was the dominant bottleneck. The global mask then more than doubles the speedup again (from $54\times$ to $138\times$), giving FastVUS. Executing the same pipeline on the GPU compounds these gains, bringing FastVUS-GPU to $677\times$ over the original VUS.}

\subsection{Scalability on Synthetic Data}
\label{sec:scalability_synthetic}

\begin{figure}
    \centering
    \includegraphics[width=\linewidth]{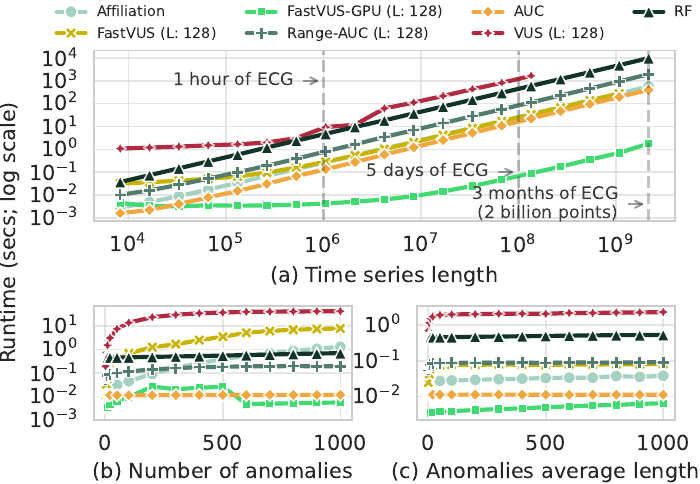}
    \Description{Runtime vs. time series characteristics.}
    \caption{Runtime versus key characteristics on synthetic time series (in (a) FastVUS and Affiliation are superposed, and ECG stands for Electrocardiogram).}
    \label{fig:exec_time_per_ts_char_syn}
\end{figure}

We next evaluate how runtime scales with key time-series characteristics using synthetic data. Figure~\ref{fig:exec_time_per_ts_char_syn}a shows runtime as a function of series length. VUS is the slowest and fails beyond roughly 100 million points, whereas FastVUS-GPU is the fastest, processing the longest series tested ($2^{31}$ points, over 2 billion samples) in \sylli{under two seconds}. Moreover, its runtime remains nearly constant up to one million points, explaining the low variance in Figure~\ref{fig:intro_runtime}. Figure~\ref{fig:exec_time_per_ts_char_syn}b examines the effect of the number of anomalies. Most methods exhibit an initial increase followed by a plateau, with VUS showing the strongest growth (from \sylli{0.2 seconds to over 40 seconds}). FastVUS follows a similar trend, reaching \sylli{8 seconds for the largest anomaly count}, while Affiliation exhibits comparable behavior at roughly an order of magnitude lower runtime. In contrast, RF, Range-AUC, and AUC are largely unaffected by the number of anomalies. FastVUS-GPU remains the fastest throughout, requiring \sylli{6 milliseconds even for the largest anomaly count}. Finally, Figure~\ref{fig:exec_time_per_ts_char_syn}c shows that anomaly length has a smaller impact on runtime than either series length or anomaly count. VUS remains the slowest, followed by RF, Range-AUC, and FastVUS, while Affiliation, AUC, and FastVUS-GPU are the least affected. Across all settings, FastVUS-GPU achieves the lowest runtime, remaining below 10 milliseconds per series even for long anomalies. \sylli{For space reasons, the runtime of VUS, FastVUS, and FastVUS-GPU as a function of the maximum buffer length, from $0$ to $512$, is reported in our repository~\cite{ffvusRepo}.}

\subsection{Practicality: \textit{Is Evaluation the Bottleneck?}}
\label{sec:practicality}

\begin{figure}[t]
    \centering
    \begin{subfigure}[b]{0.575\linewidth}
        \centering
        \caption{Evaluation vs. detection cost}
        \includegraphics[width=\linewidth]{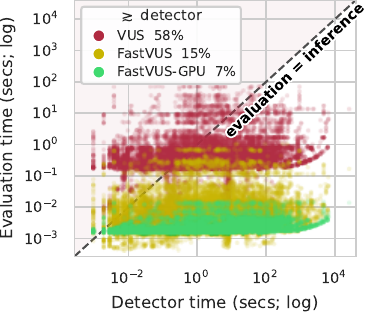}
        \label{fig:runtime_vs_detector}
    \end{subfigure}
    \hfill
    \begin{subfigure}[b]{0.405\linewidth}
        \centering
        \caption{Deep-learning pipeline}
        \includegraphics[width=\linewidth]{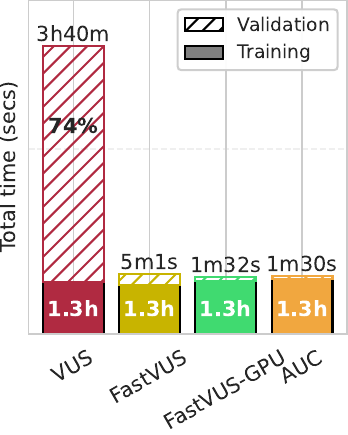}
        \label{fig:toy_example}
    \end{subfigure}
    \vspace{-0.7cm}
    \Description{Practicality of FastVUS: evaluation-versus-detector time and a deep-learning pipeline runtime breakdown.}
    \caption{\sylli{Practicality of FastVUS. (a) Measure versus detector inference time on TSB-AD, one point per series and detector pair. (b) Training (solid) and validation (hatched) time of a 100-epoch model-selection pipeline, per measure. Total validation time above each bar.}}
    \label{fig:practicality}
\end{figure}

\newpart{\noindent\textbf{Evaluation versus detection cost.} A recurring question is whether evaluating VUS is ever the bottleneck, since the detectors themselves can be expensive to run. \sylli{Figure~\ref{fig:runtime_vs_detector} answers this by plotting the measure evaluation time against the detector inference time, for every time series and detector pair in TSB-AD. The diagonal marks where evaluation costs as much as detection, and above it evaluation is the bottleneck.} The original VUS is within the same order of magnitude or slower than the detector in 58\% of runs. FastVUS and FastVUS-GPU are comparably expensive in only 15\% and 7\% of runs, reducing evaluation to \sylli{a small fraction} of the pipeline. VUS was therefore a genuine bottleneck, and FastVUS removes it.}

\noindent\textbf{Integration in deep learning pipelines.} Deep learning approaches for time series anomaly detection have grown rapidly~\cite{zamanzadeh2024deep}, making efficient evaluation measures essential\sylli{.} To demonstrate the benefits of FastVUS-GPU in this context, we designed a model selection pipeline where a 3-layer MLP is trained for 100 epochs on 1{,}400 time series from TSB-UAD~\cite{paparrizosTSB2022} (1{,}153 for training, 247 for validation). After each epoch, we evaluate the model using AUC (via \textit{torchmetrics} on GPU), VUS, FastVUS, and FastVUS-GPU. We set the buffer length to 512, which, on an Electrocardiogram (ECG) signal sampled at 256~Hz, corresponds to a 2-second tolerance window around each anomaly. Figure~\ref{fig:toy_example} presents the runtime breakdown for each measure. VUS dominates the pipeline, consuming $132$ seconds per epoch on validation alone, a $282\%$ overhead relative to training ($74\%$ of the total pipeline time). In contrast, FastVUS reduces validation to $3$ seconds per epoch ($6.5\%$ overhead), a $43.9\times$ speedup, and FastVUS-GPU to under one second per epoch ($2.0\%$ overhead), a $142.3\times$ speedup that matches AUC ($145.7\times$). Consequently, both FastVUS variants complete the 100-epoch run in about $1.4$ hours, same as AUC, compared to about $5$ hours with VUS.


\section{AutoVUS Experimental Evaluation}
\label{sec:case}

\newpart{In this section, we evaluate how AutoVUS changes \sylli{TSAD} assessment\sylli{. We first examine two cases from the TSB-AD benchmark (Section~\ref{ssec:benchmark-case}), then use controlled single-anomaly and multi-anomaly cases to isolate the effects of adaptive buffering (Sections~\ref{ssec:controlled-case} and~\ref{ssec:case-multi}), and finally compare the detector rankings of VUS and AutoVUS over the whole TSB-AD benchmark (Section~\ref{ssec:benchmark-wide}).}}

\subsection{\sylli{Real Series Case Study}}
\label{ssec:benchmark-case}

\begin{figure}
    \centering
    \includegraphics[width=\linewidth]{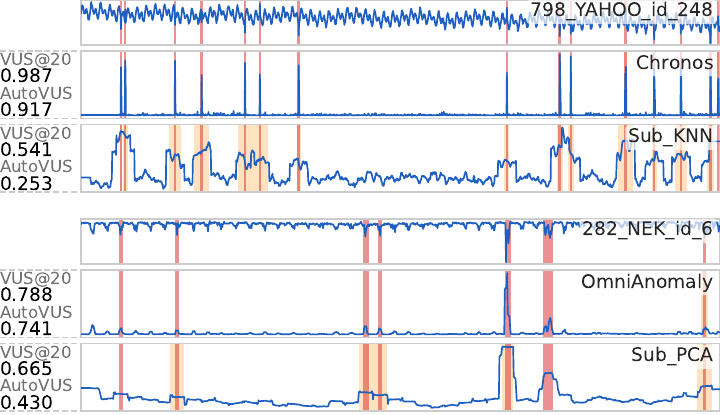}
    \caption{\sylli{Two series and the anomaly scores of two detectors each. Red marks the labeled anomalies and orange the buffers AutoVUS selects. VUS@20 is VUS-PR with $L=20$.}}
    \label{fig:tsbad-case-study}
\end{figure}

\newpart{\sylli{Figure~\ref{fig:tsbad-case-study} illustrates the effect of AutoVUS on two TSB-AD time series, 798\_YAHOO\_id\_248 and 282\_NEK\_id\_6, with two detectors each, as implemented in TSB-AD~\cite{liu2024elephant}. On the first series, Chronos responds sharply on the labeled anomalies, whereas Sub\_KNN produces broad score variations that do not consistently align with them. On the second series, OmniAnomaly peaks on the labeled anomalies, whereas Sub\_PCA rises in broader plateaus that extend around them. VUS@20 already separates the two pairs, ranking Chronos ($0.987$) above Sub\_KNN ($0.541$) and OmniAnomaly ($0.788$) above Sub\_PCA ($0.665$). AutoVUS preserves this ordering and widens the gap, as Chronos and OmniAnomaly remain high at $0.917$ and $0.741$, while Sub\_KNN and Sub\_PCA decrease to $0.253$ and $0.430$. This shows that AutoVUS does not reduce all VUS values uniformly. Instead, it preserves high scores when the detector response is supported by the anomalous regions, while reducing scores whose apparent performance relies on broad or weakly localized buffer-induced reward.}}

\subsection{Controlled Case Study}
\label{ssec:controlled-case}

\begin{figure}
    \centering
    \includegraphics[width=\linewidth]{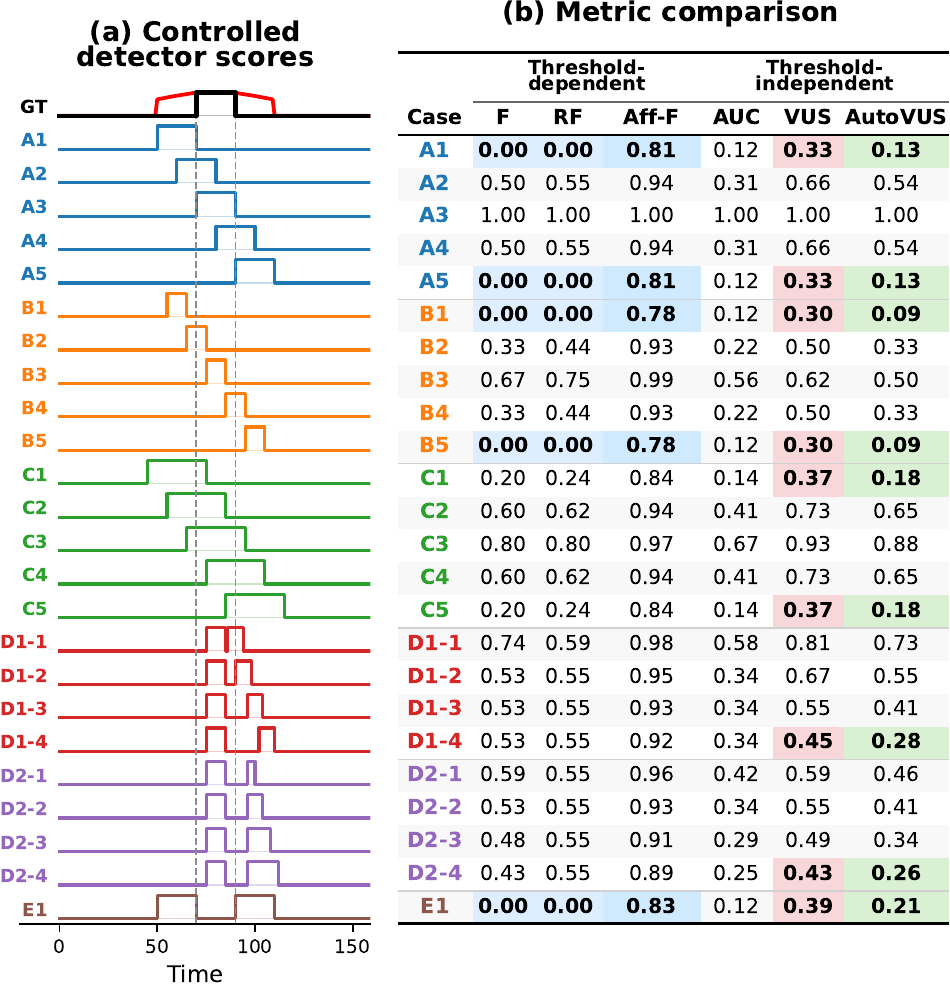}
    \caption{Controlled case study comparing evaluation measures under temporal shift, length mismatch, and extra detections. (a) Ground-truth anomaly and synthetic detector scores. (b) Threshold-dependent and threshold-independent measure values. Colored cells highlight values discussed.}
    \label{fig:controlled_cases}
\end{figure}

\newpart{We use a controlled case study to isolate what adaptive buffering changes in the evaluation. Figure~\ref{fig:controlled_cases}\sylli{a} shows a single ground-truth anomaly \sylli{on top} and several synthetic detector scores \sylli{below it}, together with representative threshold-dependent and threshold-independent measures \sylli{(Figure~\ref{fig:controlled_cases}b)}.
The red curve on the ground-truth row shows the maximum soft buffer used by VUS. Buffer lengths and maximum buffer lengths are matched to the anomaly length, so any difference between VUS and AutoVUS comes from how the buffer reward is selected and normalized, rather than from different tolerance scales. \sylli{As AutoVUS is chance-adjusted, its values are not directly comparable to those of VUS, so the focus of this study is the spread of values within each measure across the cases.}

The first \sylli{three} groups test standard temporal-tolerance behavior. Group A keeps the predicted length equal to the anomaly length and shifts it from early to late\sylli{,} Group B uses shorter predictions\sylli{,} and Group C \sylli{longer ones}. These cases highlight the different roles of the threshold-dependent measures. F has no temporal adjustment and assigns zero score to detections outside the labeled interval, even when close to the anomaly boundary. RF is also conservative when there is no overlap with the labeled interval. Aff-F, in contrast, can remain high because it rewards temporal proximity between predicted and ground-truth events. This is illustrated in A1/A5 and B1/B5, where F and RF are zero but Aff-F remains high.
}

\newpart{
The threshold-independent measures show the main effect of adaptive buffering. The exact or nearly centered detections receive high scores under AUC, VUS, and AutoVUS, while detections farther from the anomaly receive lower scores. However, VUS remains higher than AUC when the prediction falls inside the expanded label, because the fixed buffer grants partial credit for near-boundary detections. AutoVUS preserves this meaningful temporal tolerance, but reduces the reward when the improvement is mainly explained by the larger buffered region\sylli{, which as a result widens the gap between exact and shifted detections}. This is why the highlighted A1/A5, B1/B5, and C1/C5 rows show lower AutoVUS values than VUS. \sylli{Note, however, that AutoVUS is measured on a chance-adjusted scale, so any AutoVUS value above $0$ means} that the detector obtains more reward than a random score under the same buffered evaluation. Thus, a value such as AutoVUS $=0.13$ in A1 may look small, but it still indicates a detection signal beyond chance. A near-boundary detection should receive some credit because it is temporally close to the anomaly, but less than a well-localized detection such as A3, where the detector exactly matches the anomaly.

Groups D1 and D2 introduce a more challenging setting, where each detector contains a partial correct detection inside the anomaly \sylli{and} an additional segment outside \sylli{it}. These cases test whether a measure can separate useful temporal tolerance from extra detections that receive credit only because they fall into the buffered region. VUS often remains high because the outside segment is rewarded after label expansion. AutoVUS is consistently lower in these cases. The highlighted D1-4 and D2-4 rows show that the outside detections still increase VUS, but AutoVUS is more conservative when a large part of the gain comes from buffer-induced reward.

Finally, Group E is a stress test for false credit. The detector fires on both sides of the anomaly up to the maximum buffer range, but \sylli{not} inside the true anomalous interval. F and RF \sylli{assign} zero score because \sylli{nothing overlaps} the anomaly, while Aff-F remains high due to temporal proximity. \sylli{Similarly,} AUC is low, whereas VUS increases from the buffer reward\sylli{, an inflation that AutoVUS reduces}. Overall, this case study illustrates the value of AutoVUS\sylli{, which} keeps VUS's tolerance to temporal imprecision, but makes the final score less sensitive to reward obtained only through excessive buffer overlap.}

\subsection{Event-Level Interpretability}
\label{ssec:case-multi}

\begin{figure}
    \centering
    \includegraphics[width=\linewidth]{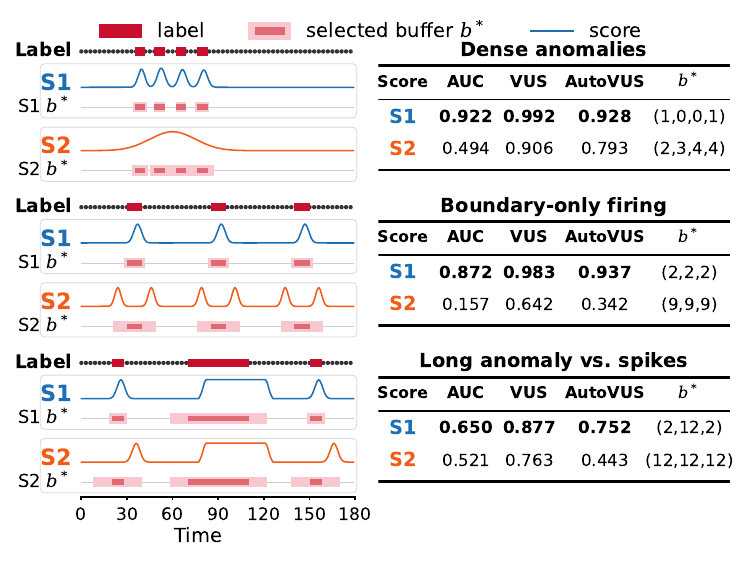}
    \caption{\sylli{
   Each case compares two detector scores under the same anomaly label sequence. The selected per-anomaly buffer vector $b^*$ is shown below each score and used to compute AutoVUS. The table reports AUC, VUS, AutoVUS, and $b^*$, all at maximum buffer length $L=20$.
    }}
    \label{fig:multi_cases}
\end{figure}

\newpart{We next use multi-anomaly controlled cases to examine the event-level behavior of AutoVUS. In addition to producing a final scalar score\sylli{, which is all the previous subsections reported}, AutoVUS selects a per-anomaly buffer vector
\(b^*=(b_1^*,\ldots,b_{N_\alpha}^*)\). These buffers are not only used in the final AutoVUS calculation, but also provide an interpretable diagnostic of detector behavior. A small \(b_k^*\) indicates that the score is well localized around the \(k\)-th anomaly, while a larger \(b_k^*\) indicates that the detector requires more temporal tolerance to match that anomaly. Therefore, \(b^*\) can reveal whether a detector responds to individual anomalous events separately or benefits from broad buffered regions.

Figure~\ref{fig:multi_cases} shows three multi-anomaly cases\sylli{, each with the anomaly label and two detector scores along with the buffers that AutoVUS selects per anomaly. Note that AutoVUS selects one buffer per anomaly and applies it on both sides, and we leave per-side selection for future work.} In the first case, the anomaly label contains four dense events. The score \(S1\) detects the four events separately, \sylli{each slightly late, so the selected buffer vector \(b^*=(1,0,0,1)\) stays small}. In contrast, \(S2\) produces one broad bridge over the dense region and treats the nearby anomalies as one wide event\sylli{, as its buffers \(b^*=(2,3,4,4)\) reveal, since its response is as high in the gaps between the events as inside them. VUS already prefers \(S1\) (\(0.992\) versus \(0.906\)), and AutoVUS widens the gap (\(0.928\) versus \(0.793\))}. This behavior is desirable because the detector that resolves the four anomalies separately should be preferred over the one that merges them into one broad response.

The second case evaluates boundary-only firing. Here, \(S1\) fires near the inside edge of each anomalous interval, while \(S2\) fires outside both sides of each anomaly. Because VUS rewards predictions that fall within the expanded buffer region, \(S2\) still receives a nontrivial VUS score \sylli{of \(0.642\)}. However, AutoVUS is more conservative, assigning \sylli{\(0.342\)} to \(S2\), compared with \sylli{\(0.937\)} for \(S1\). This shows that AutoVUS preserves temporal tolerance for near-boundary detections, but does not award excessive credit to scores that fire mostly outside the anomalies.

\sylli{The third case contains one long anomaly and two spikes. \(S1\) detects all three with a lag that grows with the anomaly length, whereas \(S2\) lags by the same amount on every anomaly. VUS prefers \(S1\) (\(0.877\) versus \(0.763\)) and AutoVUS widens the gap (\(0.752\) versus \(0.443\)). The selected buffers show why, as \(b^*=(2,12,2)\) for \(S1\) against \((12,12,12)\) for \(S2\), so the buffer adapts to each anomaly within the same series, which a single fixed buffer cannot do.}}

\sylli{\subsection{Benchmark-Wide Comparison}
\label{ssec:benchmark-wide}

\begin{figure}
    \centering
    \includegraphics[width=\linewidth]{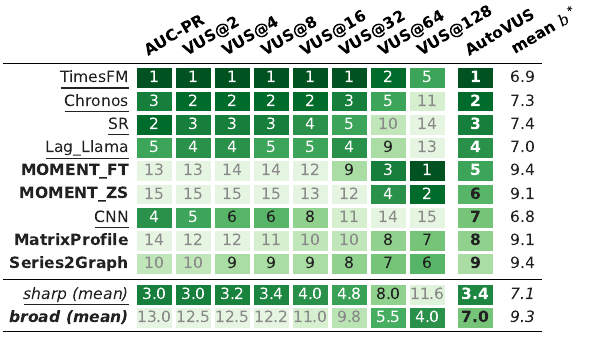}
    \caption{Rank of the nine best detectors under AutoVUS on TSB-AD. Underlined names select no buffer on most anomalies (sharp), bold names select one on most (broad). \sylli{Last two rows, mean rank per group. Last column, mean selected buffer $b^*$.}}
    \label{fig:rank_heatmap}
\end{figure}

Finally, in Figure~\ref{fig:rank_heatmap} we compare the global rankings of the detectors over TSB-AD under AUC, VUS with maximum buffer length $L$ from $2$ to $128$, and AutoVUS at $L = 128$, which takes $25$ minutes to compute over the whole benchmark. We split the detectors into two groups, the sharp group (underlined), where AutoVUS selects no buffer ($b^* = 0$) on $85\%$ of the anomalies on average, and the broad group, where it selects a buffer on $68\%$ of them. The mean ranks of the two groups show that the VUS rankings vary with $L$, since longer buffers reward the broad detectors more, whereas AutoVUS finds a balance between the two. The mean rank of the sharp group under VUS goes from $3.0$ at $L = 2$ to $11.6$ at $L = 128$, while that of the broad group goes from $12.5$ to $4.0$, overtaking the sharp group. AutoVUS ranks the sharp group first, at a mean rank of $3.4$, and the broad group second, at $7.0$.

Overall, the AutoVUS ranking stays closer to the VUS ranking than the VUS rankings at different $L$ stay to each other. The Kendall $\tau$ between the global rankings of AutoVUS and VUS@128 is $0.72$, against $0.54$ between VUS@128 and VUS@16. Finally, AutoVUS agrees with VUS on the best detector in $78\%$ of the series on average over $L$, and in the remaining $22\%$ the best detector under AutoVUS was already in the VUS top three on $70\%$ of the cases. The ranking of all $32$ detectors and further results are reported in our repository~\cite{ffvusRepo}.}


\section{Conclusion}
\label{sec:conclusion}

The Volume Under the Surface (VUS) is a time-aware measure for time series anomaly detection, but its computational cost has limited its use at scale. Moreover, VUS uses the same buffer length
for every anomaly and detector, even though the appropriate temporal tolerance may differ across cases.
In this work, we address the scalability limitation by introducing \textbf{FastVUS} and \textbf{FastVUS-GPU}\sylli{, and the buffer-length limitation by introducing \textbf{AutoVUS}. The former two} make VUS both time-aware and scalable\sylli{, computing VUS $76\times$ faster on average on TSB-AD with exactly the same scores, and $304\times$ faster with a maximum error of $5\times10^{-3}$}, \sylli{approaching} the runtime of pointwise measures such as AUC. 
\sylli{The latter is} an adaptive measure that builds on VUS by selecting a buffer for each anomaly and detector \sylli{against a random baseline}, turning VUS into a per-anomaly evaluation measure \sylli{that reduces buffer-induced score inflation}. 
\sylli{We} release FastVUS and AutoVUS as open source~\cite{ffvusRepo}\sylli{, so that the community can adopt} time-aware evaluation without \sylli{its former cost}. 

\section*{AI-Generated Content Acknowledgment}
\begin{newpartenv}
The authors used LLMs to assist with code development and with text editing for clarity. All technical content, experiments, results, and conclusions were produced, reviewed, and verified by the authors, who take full responsibility for the paper.
\end{newpartenv}

\bibliographystyle{IEEEtran}
\bibliography{sylli_bib_2025}

\end{document}